%% file: main.tex
\documentclass[conference,final,10pt,a4paper]{IEEEtran}

\usepackage[latin1]{inputenc}
\usepackage{cite}
\usepackage[pdftex]{graphicx}
 \graphicspath{{../pdf/}{../jpeg/}}
 \DeclareGraphicsExtensions{.pdf,.jpeg,.png}
\usepackage{pgf}
\usepackage{pgfplots}
\usepackage[cmex10]{amsmath}
\usepackage{amssymb}
\usepackage{array}
\usepackage{mdwmath}
\usepackage{mdwtab}
\usepackage{url}
\usepackage{fixedmultirow}
\usepackage{trfsigns}
\usepackage{fancyhdr}
\usepackage{nicefrac}
\usepackage{mathtools} 
\usepackage{algorithm}
\usepackage{algpseudocode}
\algrenewcommand\algorithmicindent{.5em}

\input{tikz-settings.tex}
\input{macros.tex}

\title{Low Complexity Decoding for Higher Order Punctured Trellis-Coded
       Modulation Over Intersymbol Interference Channels}
\author{
 \IEEEauthorblockN{Fabian~Schuh and
                   Johannes~B.~Huber}
 \IEEEauthorblockA{Institute for Information Transmission,
                   Friedrich-Alexander-Universit\"at Erlangen-N\"urnberg, Germany\\ 
                   mail: \texttt{\{schuh,\,huber\}@LNT.de}}%
}

\begin{document}
\sloppy
\maketitle
\begin{abstract}
Trellis-coded modulation (TCM) is a power and bandwidth efficient digital
transmission scheme which offers very low structural delay of the data stream.
Classical TCM uses a signal constellation of twice the cardinality compared to
an uncoded transmission with one bit of redundancy per PAM symbol, \ie,
application of codes with rates $\frac{n-1}{n}$ when $2^{n}$ denotes the
cardinality of the signal constellation.
Recently published work allows rate adjustment for TCM by means of
puncturing the convolutional code (CC) on which a TCM scheme is based on.
In this paper it is shown how punctured TCM-signals transmitted over
intersymbol interference (ISI) channels can favorably be decoded. Significant
complexity reductions at only minor performance loss can be achieved by means
of reduced state sequence estimation.
\end{abstract}
\begin{IEEEkeywords}
trellis-coded modulation (TCM);
punctured convolutional codes;
Viterbi-Algorithm (VA);
reduced state sequence estimation (RSSE);
intersymbol interference (ISI);
\end{IEEEkeywords}
\IEEEpeerreviewmaketitle
\section{Introduction                                             } \input{content_introduction               }\vspace*{0ex}
\section{System Model                                             } \input{content_systemmodel                }\vspace*{0ex}
\section{Trellis-based Decoding                                   } \input{content_ptcm-isi-uncoded-trellis   }\vspace*{-1ex}
\section{Reduced-State Sequence Estimation                        } \input{content_ptcm-isi-uncoded-rsse      }
\subsection{State Reduction Techniques (PAM without channel code) } \input{content_md-rsse-techniques         }

\subsection{Implementation Issues                                 } \input{content_ptcm-isi-uncoded-rsse-impl }\vspace*{-2ex}
\subsection{Numerical Results                                     } \input{content_ptcm-isi-uncoded-rsse-num  }\vspace*{-2ex}
\section{Conclusion                                               } \input{content_conclusion                 } 


\vspace{-3ex}
\enlargethispage{3ex}
\bibliographystyle{IEEEtran}
\bibliography{IEEEabrv,main,mine}

\end{document}

%% file: tikz-settings.tex
\graphicspath{{graphics/}}
\graphicspath{{../Bilder/}}

\usetikzlibrary{calc,trees,positioning,arrows,chains,shapes.geometric,%
 decorations.pathreplacing,decorations.pathmorphing,shapes,%
 matrix,shapes.symbols,plotmarks,decorations.markings,shadows,fit,patterns}
\usetikzlibrary{spy,backgrounds}
\usepgfplotslibrary{groupplots}

\tikzset{every picture/.style={>=latex}} 
\pgfplotsset{compat=1.3}
\pgfplotsset{filter discard warning=false} 
\pgfplotsset{every axis label/.append style={font=\small}}
\pgfplotsset{every tick label/.append style={font=\footnotesize}}
\pgfsetplotmarksize{2pt}

\tikzstyle{EMTY}       =  [ fill=white, ]
\tikzstyle{STATE}      =  [ fill=black!40!white, ]
\tikzstyle{INPUT}      =  [ fill=black!20!white, ]
\tikzstyle{UNCOD1}     =  [ pattern=crosshatch ]
\tikzstyle{UNCOD2}     =  [ opacity=.4,fill=yellow!50!orange ]
\tikzstyle{UNCOD3}     =  [ opacity=.4,fill=yellow!20!orange ]
\tikzstyle{UNCOD1L}    =  [ draw=yellow!80!orange, ]
\tikzstyle{UNCOD2L}    =  [ draw=yellow!50!orange ]
\tikzstyle{UNCOD3L}    =  [ draw=yellow!20!orange ]
\tikzstyle{EXTEND}     =  [ pattern=north east lines, ]
\tikzstyle{TRRSSE}     =  [ pattern=north west lines, ]
\tikzstyle{XSB}        =  [ thick, |-|, shorten <=3pt, shorten >=3pt ]

\tikzset{orientation/.is choice,
    orientation/lr/.style={anchor=west,right=1},
    orientation/lr2/.style={anchor=west,right=2},
    orientation/lrd/.style={anchor=west,below=1},
    orientation/lrd2/.style={anchor=west,below=2},
    orientation/rl/.style={anchor=east,left=1},
    orientation/rl2/.style={anchor=east,left=2},
    orientation/ud/.style={anchor=north,below=1},
    orientation/du/.style={anchor=south,above=1},
    orientation/rld/.style={anchor=east,below=1},
    orientation/rld2/.style={anchor=east,below=2},
}
\tikzstyle{scare} = [
]
\tikzstyle{syslinear} = [
 drop shadow={shadow xshift=.6mm,
              shadow yshift=-.6mm},
 fill=white,
 anchor=west,
 rectangle,
 draw=black,
 minimum height=5mm,
 minimum width=5mm,
 inner xsep=0.5em
]

\tikzstyle{sysnonlinear} = [
 drop shadow={shadow xshift=.8mm,
              shadow yshift=-.8mm},
 fill=white,
 double,
 anchor=west,
 rectangle,
 draw=black,
 minimum height=5mm,
 minimum width=5mm,
 inner xsep=0.5em
]

\tikzstyle{syssource} = [
 anchor=west,
 ellipse,
 draw=black,
 minimum height=1.5ex,
 minimum width=1.5em,
 inner xsep=0.5em
]

\tikzstyle{syssink} = [
 anchor=west,
 ellipse,
 draw=black,
 minimum height=1.5ex,
 minimum width=1.5em,
 inner xsep=0.5em
]

\tikzstyle{syssplit} = [
 fill=black,
 draw=black,
]

\tikzstyle{sysadd} = [
 draw,circle,inner sep=-1pt,
]

\tikzstyle{sysmul} = [
 draw,circle,inner sep=-1pt,
]

\definecolor{MyHSBGreen}{hsb}{0.34065,1,0.91}
\pgfplotscreateplotcyclelist{colors4}{%
 {black!100!white},
 {black!75!white},
 {black!50!white},
 {black!25!white},
}
\pgfplotscreateplotcyclelist{colors5}{%
 {black!100!white},
 {black!75!white},
 {black!50!white},
 {black!25!white},
 {densely dashed, black!100!white},
 {densely dashed, black!75!white},
}
\pgfplotscreateplotcyclelist{colors10}{%
 {black!100!white},
 {black!75!white},
 {black!50!white},
 {black!25!white},
 {densely dashed, black!100!white},
 {densely dashed, black!75!white},
 {densely dashed, black!50!white},
 {densely dashed, black!25!white},
 {densely dotted, black!100!white},
 {densely dotted, black!75!white},
 {densely dotted, black!50!white},
}

\pgfplotscreateplotcyclelist{colorsBERMDDFSE}{%
                 every mark/.append style={fill=black},mark=o\\%
                 every mark/.append style={fill=black},mark=star\\%
                 every mark/.append style={fill=black},mark=triangle\\%
                 every mark/.append style={scale=.7,fill=black},mark=square\\%
                 every mark/.append style={fill=black},mark=diamond\\%
  densely dashed,every mark/.append style={solid},mark=pentagon*\\%
  densely dashed,every mark/.append style={solid},mark=*\\%
  densely dashed,every mark/.append style={solid},mark=triangle*\\%
  densely dashed,every mark/.append style={scale=.7,solid},mark=square*\\%
  densely dashed,every mark/.append style={solid},mark=diamond*\\%
  densely dashed,every mark/.append style={solid},mark=star*\\%
}
\pgfplotscreateplotcyclelist{colors1Empty4Full}{%
                 every mark/.append style={fill=black},mark=o\\%
  densely dashed,every mark/.append style={solid},mark=*\\%
  densely dashed,every mark/.append style={solid},mark=triangle*\\%
  densely dashed,every mark/.append style={scale=.7,solid},mark=square*\\%
  densely dashed,every mark/.append style={solid},mark=diamond*\\%
}

\pgfplotscreateplotcyclelist{colors4Empty4Full}{%
                 every mark/.append style={fill=black},mark=o\\%
                 every mark/.append style={fill=black},mark=triangle\\%
                 every mark/.append style={scale=.7,fill=black},mark=square\\%
                 every mark/.append style={fill=black},mark=diamond\\%
  densely dashed,every mark/.append style={solid},mark=*\\%
  densely dashed,every mark/.append style={solid},mark=triangle*\\%
  densely dashed,every mark/.append style={scale=.7,solid},mark=square*\\%
  densely dashed,every mark/.append style={solid},mark=diamond*\\%
}
\pgfplotscreateplotcyclelist{colors6Empty4Full}{%
                 every mark/.append style={fill=black},mark=o\\%
                 every mark/.append style={fill=black},mark=star\\%
                 every mark/.append style={fill=black},mark=triangle\\%
                 every mark/.append style={scale=.7,fill=black},mark=square\\%
                 every mark/.append style={fill=black},mark=diamond\\%
                 every mark/.append style={fill=black},mark=pentagon\\%
  densely dashed,every mark/.append style={solid},mark=*\\%
  densely dashed,every mark/.append style={solid},mark=triangle*\\%
  densely dashed,every mark/.append style={scale=.7,solid},mark=square*\\%
  densely dashed,every mark/.append style={solid},mark=diamond*\\%
}
\pgfplotscreateplotcyclelist{colors8Empty4Full}{%
                 every mark/.append style={fill=black},mark=o\\%
                 every mark/.append style={fill=black},mark=star\\%
                 every mark/.append style={fill=black},mark=triangle\\%
                 every mark/.append style={scale=.7,fill=black},mark=square\\%
                 every mark/.append style={fill=black},mark=diamond\\%
                 every mark/.append style={fill=black},mark=pentagon\\%
                 every mark/.append style={fill=black},mark=oplus\\%
                 every mark/.append style={fill=black},mark=otimes\\%
  densely dashed,every mark/.append style={solid},mark=*\\%
  densely dashed,every mark/.append style={solid},mark=triangle*\\%
  densely dashed,every mark/.append style={scale=.7,solid},mark=square*\\%
  densely dashed,every mark/.append style={solid},mark=diamond*\\%
}

\pgfplotscreateplotcyclelist{colors}{%
                 every mark/.append style={fill=black},mark=o\\%
                 every mark/.append style={fill=black},mark=star\\%
                 every mark/.append style={fill=black},mark=triangle\\%
                 every mark/.append style={scale=.7,fill=black},mark=square\\%
                 every mark/.append style={fill=black},mark=diamond\\%
                 every mark/.append style={fill=black},mark=pentagon\\%
                 every mark/.append style={fill=black},mark=oplus\\%
                 every mark/.append style={fill=black},mark=otimes\\%
                 every mark/.append style={fill=black},mark=asterisk\\%
                 every mark/.append style={fill=black},mark=+\\%
                 every mark/.append style={fill=black},mark=-\\%
                 every mark/.append style={fill=black},mark=|\\%
                 every mark/.append style={fill=black},mark=x\\%
                 every mark/.append style={fill=black},mark=*\\%
                 every mark/.append style={fill=black},mark=triangle*\\%
                 every mark/.append style={scale=.7,fill=black},mark=square*\\%
                 every mark/.append style={fill=black},mark=diamond*\\%
                 every mark/.append style={fill=black},mark=star*\\%
  densely dashed,every mark/.append style={solid,fill=gray},mark=o\\%
  densely dashed,every mark/.append style={solid,fill=gray},mark=triangle\\%
  densely dashed,every mark/.append style={scale=.7,solid,fill=gray},mark=square\\%
  densely dashed,every mark/.append style={solid,fill=gray},mark=diamond\\%
  densely dashed,every mark/.append style={solid,fill=gray},mark=star\\%
  densely dashed,every mark/.append style={solid,fill=gray},mark=pentagon*\\%
  densely dashed,every mark/.append style={solid,fill=gray},mark=*\\%
  densely dashed,every mark/.append style={solid,fill=gray},mark=triangle*\\%
  densely dashed,every mark/.append style={scale=.7,solid,fill=gray},mark=square*\\%
  densely dashed,every mark/.append style={solid,fill=gray},mark=diamond*\\%
  densely dashed,every mark/.append style={solid,fill=gray},mark=star*\\%
  mark=text,text mark=a\\%
  mark=text,text mark=b\\%
  mark=text,text mark=c\\%
  mark=text,text mark=d\\%
  mark=text,text mark=e\\%
  mark=text,text mark=f\\%
  mark=text,text mark=g\\%
}

\pgfdeclarelayer{background layer}
\pgfdeclarelayer{foreground layer}
\pgfsetlayers{background layer,main,foreground layer}

%% file: macros.tex
\newcommand{\ie}{\emph{i.e.}}
\newcommand{\eg}{\emph{e.g.}}
\newcommand{\cf}{\emph{cf.}}

\newcommand{\lvec}[1]{\ensuremath{\mathrm{{#1}}}}  
\newcommand{\gvec}[1]{\ensuremath{{#1}}}       

%% file: content_introduction.tex
Ungerboeck's trellis-coded modulation (TCM)~\cite{1056454,UngerboeckTCM87} is
an attractive digital transmission scheme when very low structural delay of the
data stream is desired. Low structural latency is ensured by the use of
convolutional codes instead of block codes (\cf~\cite{LIT_tr_com_2009_hehn})
and the dispense with interleaving (as opposed to convolutionally
bit-interleaved coded modulation~\cite{141453}). By expanding a constellation
from $2^{n-1}$ to $2^{n}$ signal points and employing a rate-$\frac{n-1}{n}$
convolutional encoder one can improve the robustness of the transmission
against noise by up to $6\,$dB without any further costs besides computational
effort~\cite{1056454}.

A recently published paper proposes to perform rate adjustment for TCM by means
of puncturing the convolutional code (CC) on which a TCM scheme is based on.
There, metric computations and trellis diagram become
time-variant~\cite{IZS14,PTCMUnderReviewICC14}.

Here, transmission over an ISI channel is considered which requires to apply a
TCM-ISI super trellis for optimum decoding. As a result, the Viterbi algorithm
(VA) has to be extended in order to handle both coded and uncoded bits in an
optimal way (Please notice, that usual iterative equalization/decoding here is
not possible due to lack of interleaver.).
%

%% file: content_systemmodel.tex
\begin{figure}[ht]
 \begin{center}\vspace*{-2ex}
  \begin{tikzpicture}[>=latex,x=1em,y=4ex,font=\footnotesize,inner sep=0.3em,
                      node distance=10mm and 4mm]
   \node[anchor=east] (in) {$\mathbf{u}_{\mathrm{c}}[l]$};
   \node[syslinear,xshift=5mm,anchor=west,at=(in.east)] (Encoder) {$\mathcal{C}$};
   \draw[o->] (in) -- (Encoder);
   \node[syslinear,right=10mm,anchor=south west,at=(Encoder.south west),minimum width=7mm] (Punct)    {$\mathcal{P}$};
   \node[syslinear,right=5mm, anchor=south west,at=(Punct.south east),minimum height=6.4ex](Labeling) {$\mathcal{L}$};
   \node[syslinear,right=7mm, anchor=west,at=(Labeling.east)]                              (Mapper)   {$\mathcal{A}$};
   \node[syslinear,right=7mm, anchor=west,at=(Mapper.east)]                                (Channel)  {$h[k]$};
   \draw[->]  ($(Encoder.east)+(0,.8ex)$) -- ($(Punct.west)+(0,.8ex)$);
   \draw[->]  ($(Encoder.east)-(0,.8ex)$) -- ($(Punct.west)-(0,.8ex)$);
   \path      ($(Punct.west)+(-15mm,3.2ex)$) node[left] {$\mathbf{u}_{\mathrm{u}}[k]$} ++ (7.5mm,0) node {\tiny{$\dots$}} ++ (6mm,0) node[coordinate] (dots) {};
   \draw[o->] ($(Punct.west)+(-15mm,2.4ex)$) -- ++(27.0mm,0);
   \draw[o->] ($(Punct.west)+(-15mm,4.0ex)$) -- ++(27.0mm,0);
   \draw[->]  ($(Punct.east)+(0,.8ex)$) -- ++(5mm,0);
   \draw[->]  ($(Punct.east)-(0,.8ex)$) -- ++(5mm,0);
   \draw (Punct.east) ++(2.3mm,0) ellipse [x radius=2pt,y radius=8pt] node[below=1.6ex] {$\mathbf{c}[k]$};
   \draw[->]  (Labeling) -- node[pos=.5,above] {$\ell[k]$} (Mapper) -- (Channel);
   \draw[very thin] (dots) +(240:3mm) --        ++(60:3mm) node[above left,font=\tiny,inner sep=1pt] {$n_\mathrm{u}$};
   \draw[very thin] (Punct.west) ++(-3mm,0)     +(240:3mm) node[below right,font=\tiny,inner sep=1pt] {$n_\mathrm{c}$} -- ++(60:3mm);
   \draw[very thin] (Encoder.west) ++(-2mm,0)   +(240:3mm) node[below,font=\tiny,inner sep=1pt] {$n_\mathrm{c}-1$} -- ++(60:3mm);
   \draw[-o] (Channel.east) -- ++(5mm,0) node[right] {$a[k]$};
  \end{tikzpicture}\vspace*{-4ex}
 \end{center}
 \caption{System model for punctured trellis-coded modulation (P-TCM) with $n_\mathrm{u}=2$ and $n_\mathrm{c}=2$.}
 \label{fig:p-tcm:sysmodel}
 \vspace*{-3ex}
\end{figure}
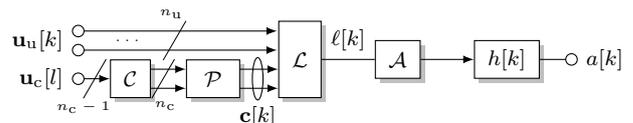

This paper deals with convolutionally encoded pulse-amplitude modulated (PAM)
transmission as depicted in Fig.~\ref{fig:p-tcm:sysmodel}. Here, the term PAM
is used for complex-valued signal constellations as well including
amplitude-shift keying (ASK), phase-shift keying (PSK) or quadrature-amplitude
modulation (QAM)
A binary data sequence $\langle u\rangle$ is split into $n_\mathrm{u}$ parallel
uncoded sequences $\mathbf{u}_\mathrm{u}[k]$ and $n_\mathrm{c}-1$ parallel sequences
$\mathbf{u}_\mathrm{c}[l]$ that are encoded using a
rate-$\frac{n_\mathrm{c}-1}{n_\mathrm{c}}$ binary convolutional encoder
$\mathcal{C}$ with generator polynomials $g_{ij}(D),\; 1 \leq i \leq
n_\mathrm{c}; 1 \leq j \leq n_\mathrm{c}-1$ with max. degree $\nu$, delay
operator $D$, $n_\mathrm{c}-1$ parallel binary-input symbols and $n_\mathrm{c}$
parallel output symbols at each time instant.

At each output of the encoder, the symbols traverse through a puncturing system
with puncturing scheme $\mathbf{P} = [P_{ij}], P_{ij}\in\left\{ 0,\,1 \right\};\;1<i\leq
n_\mathrm{c};\;1<j<\Omega$ and period $\Omega$. For each
($n_\mathrm{c}-1$)-tuple of encoder input symbols the puncturing scheme
cyclically advances by one step. Where $P_{ij}$ is zero, the current symbol at
the output is discarded, accordingly.
The punctured $n_\mathrm{c}$-ary encoded output symbols $\mathbf{c}[k]$
together with the uncoded input symbols $\mathbf{u}_{\mathrm{u}}[l]$ form a label
$\ell[k]$ by which the corresponding signal point out of
$M=2^{n_\mathrm{u}+n_\mathrm{c}}$-ary constellation is selected.
The transmit signal traverses through a dispersive discrete-time ISI channel
modelled by a FIR-filter with memory $L$ for $L+1$ channel coefficients
$h[k],\,\,0<k<L$ (\ie, $T$-spaced sampling after the Whitened Matched
Filter~\cite{Forney1972}).

%% file: content_ptcm-isi-uncoded-trellis.tex
\begin{figure}\vspace*{0ex}
 \begin{center}
\includegraphics[width=.5\textwidth]{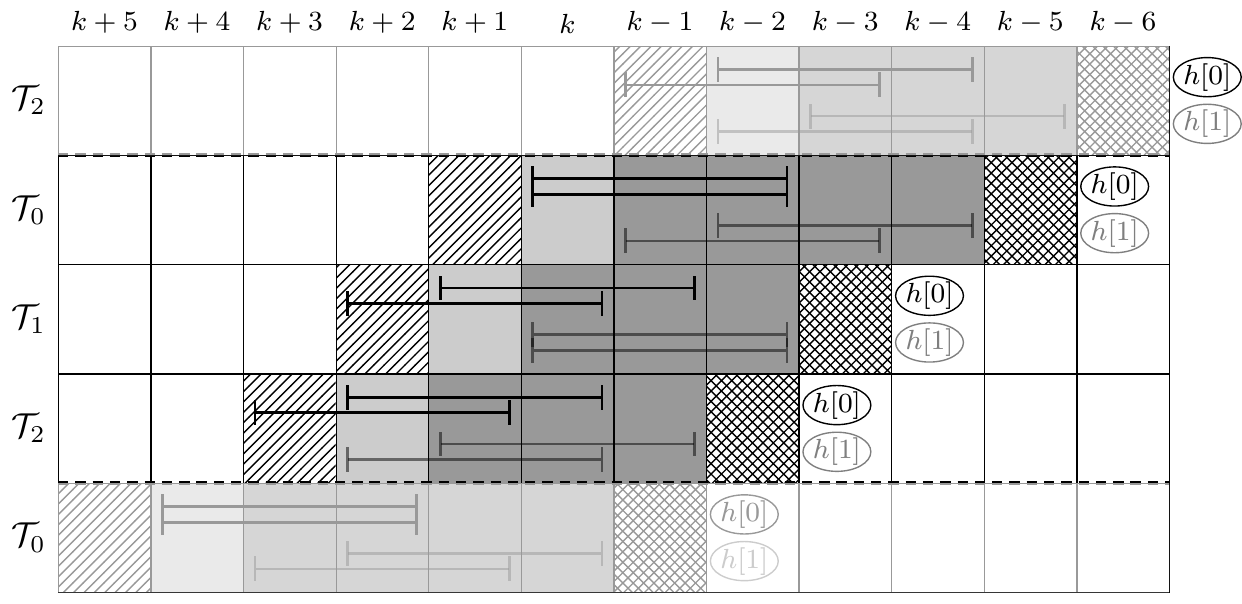}\vspace*{-4ex}
 \end{center}
 \caption{State transitions of the transmitter FSM with $R=\frac73$ and the
          relations between generator polynomials, FSM-state/input and channel
          state for a memory-$1$ ISI-channel with one uncoded bit stored in the
          states (crosshatched block).}
 \label{fig:MatchedPuncturedConvISI}
 \vspace*{-4ex}
\end{figure}

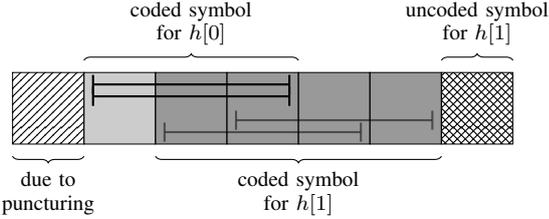
\begin{figure}
 \begin{center}
   \begin{tikzpicture}
    \tikzset { row 1 column 7/.style= { nodes= { UNCOD1 }  }  } 
    \tikzset { row 1 column 6/.style= { nodes= { STATE  }  }  } 
    \tikzset { row 1 column 5/.style= { nodes= { STATE  }  }  } 
    \tikzset { row 1 column 4/.style= { nodes= { STATE  }  }  } 
    \tikzset { row 1 column 3/.style= { nodes= { STATE  }  }  } 
    \tikzset { row 1 column 2/.style= { nodes= { INPUT  }  }  } 
    \tikzset { row 1 column 1/.style= { nodes= { EXTEND }  }  } 
    \matrix (FIFO) [matrix of nodes,
                    nodes in empty cells,
                    nodes={draw,
                           ultra thin,
                           anchor=south,
                           rectangle,
                           text width=2em,
                           minimum height=6ex,
                          },
                    ampersand replacement=\&,
                   ] {
      \&\&\&\&\&\&\\
     };
    \draw[XSB] ($(FIFO-1-2.west)+(0,2ex)$)            -- ($(FIFO-1-4.east)+(0,2ex)$);
    \draw[XSB] ($(FIFO-1-2.west)+(0,1ex)$)            -- ($(FIFO-1-4.east)+(0,1ex)$);
    \draw[XSB,opacity=.5] ($(FIFO-1-4.west)-(0,1ex)$) -- ($(FIFO-1-6.east)-(0,1ex)$);
    \draw[XSB,opacity=.5] ($(FIFO-1-3.west)-(0,2ex)$) -- ($(FIFO-1-5.east)-(0,2ex)$);
    \draw[decorate,decoration=brace] ($(FIFO-1-2.south west)-(0,1ex)$) -- node[below=.5ex,font=\footnotesize,midway] {\parbox{6em}{\centering due to\\puncturing}} ($(FIFO-1-1.south west)-(0,1ex)$);
    \draw[decorate,decoration=brace] ($(FIFO-1-2.north west)+(0,1ex)$) -- node[above=.5ex,font=\footnotesize,midway] {\parbox{6em}{\centering coded symbol for $h[0]$}} ($(FIFO-1-5.north west)+(0,1ex)$);
    \draw[decorate,decoration=brace] ($(FIFO-1-7.south west)-(0,1ex)$) -- node[below=.5ex,font=\footnotesize,midway] {\parbox{6em}{\centering coded symbol for $h[1]$}} ($(FIFO-1-3.south west)-(0,1ex)$);
    \draw[decorate,decoration=brace] ($(FIFO-1-7.north west)+(0,1ex)$) -- node[above=.5ex,font=\footnotesize,midway] {\parbox{8em}{\centering uncoded symbol for $h[1]$}} ($(FIFO-1-7.north east)+(0,1ex)$);
  \end{tikzpicture}\vspace*{-4ex}
 \end{center}
 \caption{Layout of a particular trellis state showing the individual
          components of the state ($n_\mathrm{u}=1$, $L=1$, $\nu=2$).}
 \label{fig:fifoPTCMISIUncoded}
 \vspace*{-4ex}
\end{figure}

\label{sec:trellisdecoding}
For sake of simplicity, we use for the following explanations a simple example,
\ie, a rate $\frac12$ mother code (\eg, $n_\mathrm{c}=1$) and a memory-$1$ ISI
channel. Fig.~\ref{fig:fifoPTCMISIUncoded} illustrates the extension of the
trellis states by the uncoded symbols that are stored in the ISI channel. The
two bars 
\mbox{(\!\!\tikz[baseline=-.6ex]\draw[XSB] (0,0) -- node[above,inner sep=1pt,font=\footnotesize,midway] {$g_i$} (3em,0);)}
represent the generator polynomials $g_1$ and $g_2$ which generate the MSB and
LSB of the (punctured) convolutional code (rate-$\frac12$ mother code). This
notation was introduced in~\cite{IZS14,PTCMUnderReviewICC14}. As the ISI
channel stores the last $L$ transmitted signal points, a super trellis needs to
track not only the coded output of the (punctured) convolutional encoder but
also the $L\cdot n_\mathrm{u}$ last uncoded symbols. Thus, for each uncoded
symbol and each channel tap (except for $h[0]$) an additional binary memory
element must be added to the FSM, \eg, for $n_\mathrm{u}=2$ and $L=2$ a total
of $4$ additional memory elements have to be spent. As a result, the joint
CC+ISI (super) trellis quickly becomes prohibitively large. However we will
show later that reduced-state sequence-estimation (RSSE) techniques can be
applied here, such that the computational complexity can be reduced
significantly with only minor performance loss.

We first briefly consider an algorithm to construct the trellis of this finite
state machine. In order to achieve optimal decoding we need to consider the
coded and uncoded symbols in the finite-state machine as they take part in the
memory of the CC and the ISI channel. However, as the uncoded symbols directly
propagate to the mapper the uncoded symbols immediately affect the selection of
the signal point and thus the input of the ISI channel.

To algorithmically handle the time-variant mapping we introduced~\cite{IZS14} a
set of so-called generator offsets $\mathcal{T}_i$ which describe, depending on
the puncturing scheme, modulation size, and time instant, the relations between
generator polynomials, input value, FSM state, and mapping to MSB or LSB,
respectively. For each new generator offset $\mathcal{T}_i$ a new trellis
segment arises, \eg, the number of generator offsets equals the number of
trellis segments in one trellis period.

The setup of a trellis representation of CC+ISI is defined by means of
algorithm~\ref{alg:FSMPTCMISIUncoded} in an abstract manner. The major steps
are in line~\ref{alg:line:getu} and line~\ref{alg:line:labeluncoded} which
fetches the uncoded symbols from the trellis state and appends it to the
encoding results in order to perform the labeling. Then, $L+1$ signal points
are selected and weighted with the channel coefficients in
line~\ref{alg:line:channel}.

\begin{algorithm}
 \caption{Building the FSM for P-TCM with $n_\mathrm{u}>0$ over ISI channels}
 \label{alg:FSMPTCMISIUncoded}
 \begin{algorithmic}[1]
  \Require $\lvec{h} \gets \left\{ h[0], h[1], h[2], \ldots \right\}$
  \Require $Z$ number of states
  \ForAll{$\lvec{s} : 1 \to Z$}
   \ForAll{$\lvec{u} \in \mathbb{F}_2^{n_\mathrm{c}+n_\mathrm{u}}$}
    \ForAll{$i : 1 \to \Lambda$}
     \State $\lvec{d} \gets \left\{ \lvec{u}, s_0, s_1, \ldots s_{Z-1} \right\}$
     \State $\lvec{s}^- \gets \Call{GetCurrentState}{\lvec{d}, \mathcal{T}_i}$
     \State $\lvec{s}^+ \gets \Call{GetNextState}{\lvec{d}, \mathcal{T}_i}$
     \For{$\kappa : 0 \to (\mathrm{length}(\lvec{h})-1)$}
      \State $\lvec{e}(\kappa) \gets \Call{P-Encoder}{\mathcal{T}_i, \kappa, \lvec{g}_\mathrm{oct}, \lvec{d}}$
     \EndFor
     \State $\lvec{u}_\mathrm{h} \gets \Call{GetUncodedFromState}{\lvec{u}, \mathcal{T}_i}$\label{alg:line:getu}
     \State $\displaystyle\gvec{\ell}\gets \Call{Labeling}{\left\{\lvec{u}_\mathrm{h} \quad,\quad \sum\limits_\kappa\lvec{e}(\kappa)\right\}}$\label{alg:line:labeluncoded}
     \State $\lvec{m} \gets \Call{Constellation}{\gvec{\ell}}$
     \State $h(\mathcal{T}_i,\lvec{s}^-,\lvec{u}) \gets \lvec{m}^\top \cdot \lvec{h}$\label{alg:line:channel}\Comment{hypotheses}
     \State $T(\mathcal{T}_i,\lvec{s}^-,\lvec{u}) \gets \lvec{s}^+$\Comment{next state}
    \EndFor
   \EndFor
  \EndFor
  \Function{P-Encoder}{$\mathcal{T}_i, \kappa, \lvec{g}_\mathrm{oct}, \lvec{d}$}
   \For{$i \in \left\{ \mathrm{MSB}, \mathrm{LSB} \right\}$}
    \State $\mathrm{FIFO}_i \gets$ shifts according to $\mathcal{T}_i$, and $\kappa$
    \State $e(i)\gets \mathrm{FIFO}_i^\top \cdot \lvec{g}_{i,\mathrm{dual}}$
   \EndFor
  \EndFunction
 \end{algorithmic}
\end{algorithm}

The routines \textsc{GetCurrentState} and \textsc{GetNextState} have to
evaluate the current and next state to construct the trellis as described in
detail in~\cite{IZS14,PTCMUnderReviewICC14}. Additionally, we here also have to
consider the uncoded symbols (\textsc{GetUncodedFromState}). The prototype
routines for our example
($\mathbf{P}=\left[\,(1\,1)^\top\,(0\,1)^\top\,\right]$, $\nu=2$, $L=1$) are
given in algorithm~\ref{alg:prototyefunctions} and return the content of the
delay elements on the FIFO according to
Fig.~\ref{fig:MatchedPuncturedConvISI}. Here, counting starts at the most
left elements for each generator offset $\mathcal{T}_i$.

\begin{algorithm}
 \caption{Prototype routines for the exemplary rate $\frac73$ transmission
          scheme with a memory-$2$ convolutional code and the puncturing scheme
          $\mathbf{P}=\left[\,(1\,1)^\top\,(0\,1)^\top\,\right]$ (see
          Fig.~\ref{fig:MatchedPuncturedConvISI}. Additional
          $n_\mathrm{u}=1$ uncoded bits are used to address the signal points.}
 \label{alg:prototyefunctions}
 \begin{algorithmic}[1]
  \Function{GetCurrentState}{$\lvec{d}, \mathcal{T}_i$}
      \If{$i = 0$} 
    \Return $\lvec{d}(\left\{ 3,4,5,6 \right\})$
   \ElsIf{$i = 1$} 
    \Return $\lvec{d}(\left\{ 2,3,4,5 \right\})$
   \ElsIf{$i = 2$} 
    \Return $\lvec{d}(\left\{ 2,3,4,5 \right\})$
   \EndIf
  \EndFunction
  \Function{GetNextState}{$\lvec{d}, \mathcal{T}_i$}
      \If{$i = 0$} 
    \Return $\lvec{d}(\left\{ 2,3,4 \right\})$
   \ElsIf{$i = 1$} 
    \Return $\lvec{d}(\left\{ 1,2,3,4 \right\})$
   \ElsIf{$i = 2$} 
    \Return $\lvec{d}(\left\{ 1,2,3,4 \right\})$
   \EndIf
  \EndFunction
 \end{algorithmic}
\end{algorithm}

As a result, a matrix of hypotheses $\mathbf{h}(\mathcal{T}_i)$ and a
matrix of state transitions $\mathbf{T}(\mathcal{T}_i)$ span a time-variant
trellis with $2^{(n_\mathrm{u}+n_\mathrm{c})}$ branches per state that can be
used to optimally decode punctured TCM over ISI channels by means of a modified
Viterbi algorithm. Here, $\mathcal{T}_i$ denotes the $i$-th trellis segment out
of the set of $\Lambda$ segments. Hence, $\Lambda$ is nominated as
\emph{trellis period}.
The modifications necessary for the Viterbi algorithm comprise two separate
path registers for uncoded and coded symbols. This is due to the differing
symbol phases w.r.t. the puncturing matrix, \eg, differing data speed of
$\mathbf{u}_\mathrm{u}$ and $\mathbf{u}_\mathrm{c}$ (\cf,
Fig.~\ref{fig:p-tcm:sysmodel}). The metric computations are described in the
next section.

%% file: content_ptcm-isi-uncoded-rsse.tex
\label{sec:trellisdecoding:rsse}
We will now describe the application of RSSE by recalling the basic principles
of Delayed Decision Feedback Sequence Estimation~\cite{Duel-Hallen1989} (DFSE).
We will briefly show the metric calculations before giving numerical simulation
results.
\vspace*{-1ex}

%% file: content_md-rsse-techniques.tex
\subsubsection{DFSE}
When equalizing uncoded digital PAM signaling over a discrete-time ISI channel
with $L+1$ taps using \emph{delayed decision feedback sequence estimation}
(DFSE), a trellis is constructed from the first $\tilde{L}\leq L$ taps only.
Thus, the number of states is reduced from $M^L$ to $M^{\tilde{L}}$.
For $\tilde{L}=L$ DFSE is equivalent to MLSE decoding via full-state VA. In
contrast, when $\tilde{L}=0$, the resulting trellis has a single state and,
thus, represents a decision feedback equalization (DFE). Hence, DFSE allows an
efficient way to trade between full-state VA and one-state
DFE~\cite{Duel-Hallen1989}.
The remaining $L+1-\tilde{L}$ channel taps are considered in a delayed
decision-feedback equalization (DFE) that is performed in each trellis state
using the \emph{delayed} path register of the corresponding state.

The main difference to full state equalization appears in the metric
computation for each time instant. From eq.~(\ref{eq:dsfeMetric}) it can be
seen that the state specific path register $p_\mathrm{reg}[k,\lvec{s}]$ is
delayed by $\tilde{L}$ and its elements are multiplied by the subsequent
channel coefficients $h_\mathrm{dfe}[h]$ which have not been considered in the
trellis. The branch metric $\lambda(\lvec{s},\lvec{u})$ (\eg, Euclidean
distance of the received symbol $y[k]$ to the hypotheses $h(\lvec{s},\lvec{u})$
for the state $\lvec{s}$ and symbols $\lvec{u}$) thus includes a DFE-term
$\delta$:
\begin{align}\label{eq:dsfeMetric}
 \delta                     & = \sum\limits_\kappa p_\mathrm{reg}[k-\tilde{L}+\kappa,\lvec{s}]\,\cdot\,h_\mathrm{dfse}[\kappa]\\
 \lambda(\lvec{s},\lvec{u}) & = \big| y[k] - h(\lvec{s},\lvec{u}) - \delta\big|^2\notag
\end{align}


\subsubsection{RSSE}
In RSSE, on the other hand, $Z$ arbitrary MLSE states, each with
$2^{n_\mathrm{c}+n_\mathrm{u}-1} = \frac{M}{2}$ possible branches to adjacent
states, are combined into $Z_\text{R} = \frac{Z}{2^J};\;J\in\mathbb{N}$
\emph{hyperstates}~\cite{huber1992trelliscodierung,Spinnler1995} each having
$2^J$ substates and $2^K\cdot2^J$ branches. A certain assignment of states to
hyperstates is called \textit{partitioning}~\cite{Spinnler1995}.

Instead of having $2^K$ arriving branches at each of the $Z$ MLSE states we get
a set of $2^K\cdot2^J$ branches at each of the $Z_\text{R}$ hyperstates. The
total number of available branches remains $2^K\cdot Z$. However, when using
RSSE only $2^K$ branches are possible (\ie, enabled) from each state, at a
given time instant. The availability of branches is determined by the path
register in each state, and is thus a form of decision-feedback.

The main difference of DFSE and RSSE to MLSE is, that we decide for a surviving
path prematurely resulting in a truncation of error events. A loss in Euclidean
distance appears if an error event with minimum Euclidean distance gets
truncated. Therefore the performance of RSSE strongly depends on the
partitioning of the states into hyperstates. However, survivor-decision
specifies the sub-state within a hyperstate uniquely and thus allows correct
metric calculation. Instead of exhaustively search for the optimum state
partitioning, which maximizes the intra-hyperstate
distance~\cite{Spinnler1995}, we exploit the minimum phase characteristics of
the ISI channel which is, as described above, fully integrated into our
trellis.

For a minimum phase channel impulse response the prior channel input symbols
are weighted fewer than more recent once and, thus, affect the metric less.
Hence, the intra-hyperstate distance is
maximized when states are combined with respect to elder positions in the state
number. This particular partitioning is equivalent to DFSE for ISI channels and
will be called \emph{DFSE partitioning}.
The minimum phase ISI channel is the last element to affect the received
symbols and is also fully integrated into the FSM. Thus we can apply the
\emph{DFSE partitioning} to use RSSE for punctured TCM (P-TCM) over ISI
channels.

%% file: content_ptcm-isi-uncoded-rsse-impl.tex
\vspace*{-2ex}
\begin{algorithm}
 \caption{Metric calculations for total RSSE -- $J$\textsuperscript{th} partition}
 \label{alg:MetricISIRSSE}
 \begin{algorithmic}[1]
  \State $q_\mathrm{c} \gets \log_2(\mathrm{nr.~hyperstates})$
  \ForAll{$\lvec{s} \in \mathcal{S}$}
   \ForAll{$\lvec{u} \in \mathcal{A} \cdot K$}
    \ForAll{$\kappa = 0 \to J-1$} \Comment{active branches}
     \If{$\kappa\leq L$}
      \State $\zeta[\kappa] =
      p_\mathrm{reg,uncoded}(L_\mathrm{traceback}-L+\kappa,\lvec{s})$
     \Else
      \State $\zeta[\kappa] = p_\mathrm{reg,coded}(L_\mathrm{traceback}-L-q_\mathrm{c}+\kappa,\lvec{s})$
     \EndIf
    \EndFor
    \State$\displaystyle\lambda(\gvec{\zeta},\lvec{u}) \gets\Gamma(\gvec{\zeta})  + \big| y[k] - h(\gvec{\zeta},\lvec{u})\big|^2$
   \EndFor
  \EndFor
 \end{algorithmic}
\end{algorithm}
Most importantly, in terms of implementation, we need to distinguish between
two path registers, namely one for the uncoded symbols $p_{\mathrm{rec,uncoded}}$,
that pass through the ISI channel only, and one for the coded symbols
$p_\mathrm{reg,coded}$.
Hence, when performing the feedback in RSSE both
registers need to be considered, depending on the partition depth $J$. 
Algorithm~\ref{alg:MetricISIRSSE} shows the application of both path
registers in order to select the available paths by means of an indicator
variable $\gvec{\zeta}$ which specifies the active branches by means of
decision-feedback.

%% file: content_ptcm-isi-uncoded-rsse-num.tex
\begin{figure*}[!t]
 \begin{center}
  \begin{tikzpicture}
   \begin{groupplot}[
                group style={group size=2 by 1},
                group/horizontal sep=3em,
                group/vertical sep=0em,
                width=.5\textwidth,
                height=6cm,
                enlargelimits=false,
                ymode=log,
                grid=both,
                ]
    \nextgroupplot[
                xlabel={$10\log_{10}\left(\frac{E_\text{b}}{N_0}\right)$ in dB},
                ylabel={BER},
                cycle list={
                                densely dotted,every mark/.append style={fill=black,solid},mark=o\\%
                                densely dotted,every mark/.append style={fill=black,solid},mark=star\\%
                                         solid,every mark/.append style={fill=black,solid},mark=triangle\\%
                                         solid,every mark/.append style={fill=black,solid},mark=asterisk\\%
                                         solid,every mark/.append style={fill=black,solid},mark=pentagon*,mark repeat=2,mark phase=1\\%
                                         solid,every mark/.append style={fill=black,solid},mark=diamond,mark repeat=2,mark phase=2\\%
                },
                xmin=5,xmax=20,
                ymax=1,ymin=1e-4,
                legend pos=north east,
                every axis legend/.append style={font={\tiny},nodes={right}},
               ]
    \addplot table[x index=0,y index=2]         {data_ptcm-isi-ber-1-1.csv};
    \addplot table[x index=0,y index=1]         {data_ptcm-isi-ber-2-1.csv};
    \addplot table[x index=0,y index=1]         {data_ptcm-isi-ber-4-1.csv};
    \addplot table[x index=0,y index=2]         {data_ptcm-isi-ber-4-1.csv};
    \addplot table[x index=0,y index=5]         {data_ptcm-isi-ber-4-1.csv};
    \addplot table[x index=0,y index=7]         {data_ptcm-isi-ber-4-1.csv};
    \addlegendentry{hard decision}
    \addlegendentry{soft decision}
    \addlegendentry{RSSE ($ 32$)}
    \addlegendentry{RSSE ($ 64$)}
    \addlegendentry{RSSE ($256$)}
    \addlegendentry{full-state VA ($2048$)}
    \coordinate (note) at (rel axis cs:0.05,0.05);
    \draw[thick] (axis cs:5,1e-3) -- (axis cs:25,1e-3);
    \nextgroupplot[
                xlabel={$10\log_{10}\left(\frac{E_b}{N_0}\right)$ in dB},
                ylabel={Complexity},
                log basis y=2,
                cycle list={
                            every mark/.append style={fill=black},mark=o\\%
                            every mark/.append style={fill=black},mark=star\\%
                            every mark/.append style={fill=black},mark=*\\%
                           },
                xmin=10,xmax=22,
                ymin=16,ymax=2500,
                every axis y label/.append style={yshift=-.8em},
               ]
    \addplot table[x index=0,y index=1] {data_ptcm-isi-complexity-1-1.csv} node[pos=0,pin={[pin distance=2ex]150:\footnotesize{DFSE(ISI)+VA(CC)}}] {}
                                                                           node[pos=1,pin={[pin distance=2ex]-30:\footnotesize{VA(ISI)+VA(CC)}}] {};
    \addplot table[x index=0,y index=1] {data_ptcm-isi-complexity-2-1.csv} node[pos=0,pin={[pin distance=2ex]60:\footnotesize{BCJR(ISI)+VA(CC)}}] {};
    \addplot table[x index=0,y index=1] {data_ptcm-isi-complexity-4-1.csv} node[pos=0.8,font=\footnotesize,sloped,anchor=south east] {$\leftarrow$ RSSE};
    \addplot table[x index=0,y index=1] {data_ptcm-isi-complexity-4-1.csv} node[pos=1,pin={[pin distance=2ex]230:\footnotesize{full-state VA}}] {};
    \node[draw,fill=white,anchor=south west,font={\footnotesize}] at (rel axis cs:0.02,0.02) {$\mathrm{BER} = 10^{-3}$};
   \end{groupplot}
  \end{tikzpicture}\vspace*{-4ex}
 \end{center}
 \caption{BER Performance and computational decoder complexity for a rate
          $\frac73$ transmission scheme with a memory-$2$ convolutional code
          ($[ 7,\,3]_8$) and the puncturing scheme
          $\mathbf{P}=\left[\,(1\,1)^\top\,(0\,1)^\top\,\right]$. Additional
          $n_\mathrm{u}=1$ uncoded bits are used to address the signal points
          and the ISI channel has memory $L=2$.}
 \label{fig:PTCMISIUncodedRSSEBER}\vspace*{-4ex}
\end{figure*}
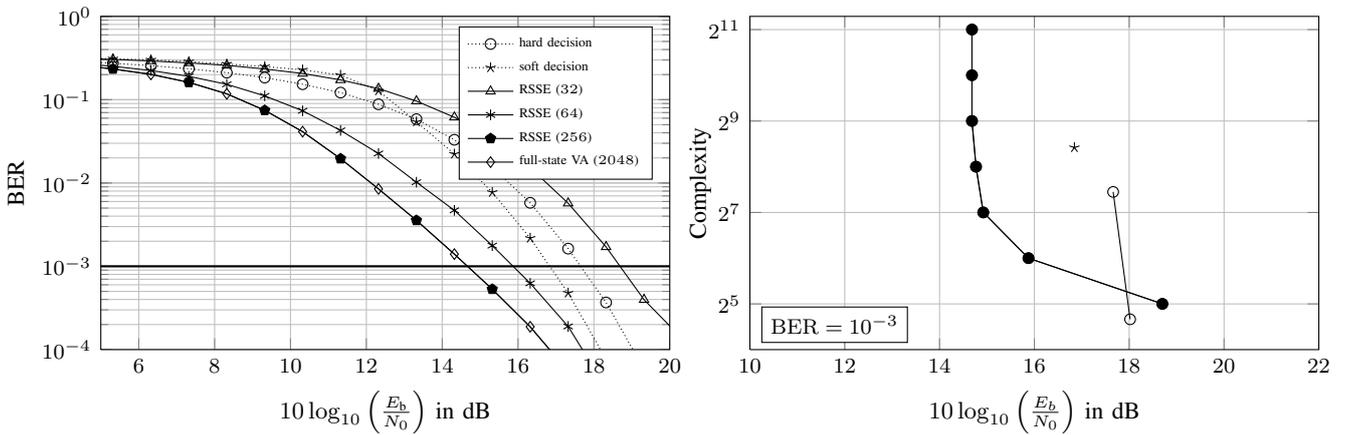

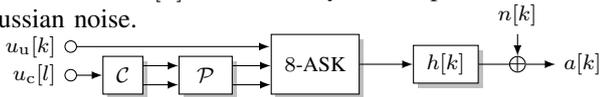
\begin{figure}[ht]
 \begin{center}\vspace*{-5ex}
  \begin{tikzpicture}[>=latex,x=1em,y=4ex,font=\footnotesize,inner sep=0.3em,
                      node distance=10mm and 4mm]
   \node[anchor=east] (in) {$u_{\mathrm{c}}[l]$};
   \node[syslinear,xshift=5mm,anchor=west,at=(in.east)] (Encoder) {$\mathcal{C}$};
   \draw[o->] (in) -- (Encoder);
   \node[syslinear,right=10mm,anchor=south west,at=(Encoder.south west),minimum width=7mm] (Punct)     {$\mathcal{P}$};
   \node[syslinear,right=5mm, anchor=south west,at=(Punct.south east),minimum height=5ex](Modulator) {$8$-ASK};
   \node[syslinear,right=7mm, anchor=west,at=(Modulator.east)]                             (Channel)   {$h[k]$};
   \draw[->]  ($(Encoder.east)+(0,.8ex)$) -- ($(Punct.west)+(0,    .8ex)$);
   \draw[->]  ($(Encoder.east)-(0,.8ex)$) -- ($(Punct.west)-(0,    .8ex)$);
   \path                  ($(Punct.west)+(-15mm,2.4ex)$) node[left] {$u_{\mathrm{u}}[k]$} ++ (7.5mm,0);
   \draw[o->]             ($(Punct.west)+(-15mm,2.4ex)$) -- ++(27.0mm,0);
   \draw[->]              ($(Punct.east)+(0,     .8ex)$) -- ++(5mm,0);
   \draw[->]              ($(Punct.east)-(0,     .8ex)$) -- ++(5mm,0);
   \draw[->]  (Modulator) -- (Channel);
   \draw node[sysadd,right=of Channel.east] (pNoise) {$+$};
   \draw[<-] (pNoise) -- ++(0,4mm) node[above] {$n[k]$};
   \draw[->] (Channel.east) -- (pNoise) -- ++(5mm,0) node[right] {$a[k]$};
   %
  \end{tikzpicture}\vspace*{-3ex}
 \end{center}
 \caption{System model for punctured trellis-coded modulation (P-TCM) with $n_\mathrm{u}=1$ and $n_\mathrm{c}=2$.}
 \label{fig:p-tcm:sysmodel8ASK}
 \vspace*{-2ex}
\end{figure}

Numerical result of a computer simulation of a transmission scheme with the
following parameters are given. The total transmission rate is $\frac73$ with a
memory-$2$ convolutional code with generator polynomials $[ 7,\,3]_8$ and the
puncturing scheme $\mathbf{P}=\left[\,(1\,1)^\top\,(0\,1)^\top\,\right]$. To
increase the transmission rate from $\frac43$ to $\frac73$ additional
$n_\mathrm{u}=1$ uncoded bits are used to address a signal points from an
$8$-ASK constellation (\cf, Fig.~\ref{fig:p-tcm:sysmodel8ASK}).
Additionally, square QAM constellations may be built from two independent ASK
constellations in in-phase and quadrature component to further increase
transmission rate without increasing the signal bandwidth. Thus, for all usual
PAM constellations P-TCM is favorably based on the one-dimensional $4$-ASK
constellation because fine tuning of the transmission rate is possible by means
of puncturing and addition of uncoded bits.
The transmit signal traverses through an ISI channel $h[\kappa]$ with memory
$L=2$ plus additive white Gaussian noise.

We use the following unit energy channel $h[\kappa]$:
\begin{align*}
 h_\mathrm{lin}[\kappa] &= \,\frac{L-\kappa+1}{L+1} \qquad \text{for } 0\leq\kappa\leq L\\
 h[\kappa] &= \frac{1}{\sqrt{\sum\limits_\gamma{|h_\mathrm{lin}[\gamma]|^2}}}\,h_\mathrm{lin}[\kappa]
\end{align*}

Figure~\ref{fig:PTCMISIUncodedRSSEBER} shows bit error rates versus
$\frac{E_\mathrm{b}}{N_0}$ from simulations. We compare
the results of our proposed approach with separated equalization and decoding
using BCJR and DFSE for soft-/hard equalization of the ISI and a full-state VA
for decoding of the CC. Note that, due to the objective of very low structural
delay of the data stream no interleaver is applicable and thus no iterative
equalization-decoding procedure is possible.
Clearly, the proposed scheme outperforms the separated approaches by several
decibels. However, the full-state trellis complexity
number\footnote{\label{fn:trelliscomplexity}The computation
trellis complexity number is defined as the number of metric calculations per
information bit.} is $2048$ and thus
significantly higher when compared to the separated approaches. Hence we reduce
the computational complexity by sacrificing optimality by reduced-state
sequence estimation (RSSE). By this, we can reduce the complexity number from
$2048$ down to $256$ without noticeable loss in performance. When further
reducing to a complexity number of $64$ the performance is still slightly
better as the soft-equalization and decoding approach, although the latter has
a complexity number of roughly $1369$.
Computational complexity\textsuperscript{\ref{fn:trelliscomplexity}} over $\frac{E_\mathrm{b}}{N_0}$ that is required to achieve a
bit error probability of less than $10^{-3}$ is shown in
Fig.~\ref{fig:p-tcm:sysmodel8ASK}. Obviously our approach allows to reduce the
complexity to $2^6$ and still outperforms the separate
soft-equalization/decoding approach.

%% file: content_conclusion.tex
\label{sec:conclusion}

It has been shown that maximum-likelihood decoding for punctured TCM can be
achieved with low computational complexity and very good performance even for
transmission over ISI channels. Thus, punctured TCM can be applied as a
low-latency transmission scheme with high spectral efficiency. An additional
benefit is the improved flexibility in transmission rate due to the flexible
choice of the puncturing scheme, in contrast to classical TCM which can only
achieve integer rates.